\documentclass[12pt]{article}

\ifx\pdfoutput\undefined
\usepackage[dvips,bookmarks]{hyperref}
\else
\usepackage{hyperref}
\fi
\hypersetup{colorlinks=false,bookmarksopen,bookmarksnumbered,citecolor=blue,
   pdfstartview=FitH}

\usepackage[dvips]{graphicx}
\usepackage{latexsym}

\usepackage{amssymb,amsfonts,amsmath}
\usepackage{graphicx} 
\usepackage{indentfirst}

 \usepackage{bbm}

\parskip=\medskipamount

\newcommand {\cD}{{\cal D}}

\newcommand {\cH}{{\cal H}}

\newcommand {\cM}{{\cal M}}
\newcommand {\cN}{{\cal N}}
\newcommand {\cO}{{\cal O}}

\def\a{\alpha}

\def\b{\beta}
\def\c{\chi}
\def\d{\delta}
\def\e{\epsilon}

\def\G{\Gamma}

\def\k{\kappa}
\def\l{\lambda}

\def\o{\omega}

\def\q{\theta}
\def\r{\rho}
\def\s{\sigma}

\def\x{\xi}
\def\z{\zeta}
\def\D{\Delta}
\def\F{\Phi}
\def\J{\Psi}
\def\L{\Lambda}
\def\O{\Omega}
\def\P{\Pi}

\def\S{\Sigma}
\def\U{\Upsilon}
\def\X{\Xi}

\newcommand{\pa}{\partial}                           
\newcommand{\hf}{\frac12}

\newcommand{\vf}{\varphi}

\newcommand{\be}{\begin{equation}}
\newcommand{\ee}{\end{equation}}
\newcommand{\bea}{\begin{eqnarray}}
\newcommand{\eea}{\end{eqnarray}}
\newcommand{\non}{\nonumber}
\newcommand{\1}{\underline{1}}
\newcommand{\2}{\underline{2}}

\def\dt#1{{\buildrel {\hbox{\LARGE .}} \over {#1}}}    

\newcommand{\bm}[1]{\mbox{\boldmath$#1$}}

\def\double #1{#1{\hbox{\kern-2pt $#1$}}}

\begin{document}
\begin{titlepage}
\begin{flushright}
UUITP-17/09\\
June, 2009\\
\end{flushright}
\vspace{5mm}

\begin{center}
{\Large \bf New extended  superconformal  sigma models
and Quaternion K\"ahler manifolds}\\ 
\end{center}

\begin{center}

{\bf
Sergei M. Kuzenko\footnote{kuzenko@cyllene.uwa.edu.au}${}^{a}$,
Ulf Lindstr\"om\footnote{ulf.lindstrom@fysast.uu.se}${}^{b}$
and Rikard von Unge\footnote{unge@physics.muni.cz}${}^{c}$ 
} \\
\vspace{5mm}

\footnotesize{
${}^{a}${\it School of Physics M013, The University of Western Australia\\
35 Stirling Highway, Crawley W.A. 6009, Australia}}  
~\\
\vspace{2mm}

\footnotesize{
${}^{b}${\it Department of Physics and Astronomy, Theoretical Physics,
Uppsala University \\ 
Box 803, SE-751 08 Uppsala, Sweden}
}
\\
\vspace{2mm}

\footnotesize{
${}^c${\it Institute for Theoretical Physics, Masaryk University, \\
61137 Brno, Czech Republic } }

\end{center}
\vspace{5mm}

\begin{abstract}
\baselineskip=14pt
Quaternion K\"ahler manifolds are known to be the target spaces for 
matter hypermultiplets coupled to $\cN=2$ supergravity. 
It is also known that there is a one-to-one correspondence 
between $4n$-dimensional quaternion K\"ahler manifolds
and those $4(n+1)$-dimensional hyperk\"ahler spaces which are the target 
spaces for rigid superconformal hypermultiplets  (such spaces are called hyperk\"ahler cones).
In this paper we present a projective-superspace construction to generate a 
hyperk\"ahler cone $\cM^{4(n+1)}_H$ of dimension $4(n+1)$
from a $2n$-dimensional real analytic K\"ahler-Hodge manifold $\cM^{2n}_K$.
The latter emerges as a maximal K\"ahler submanifold of the $4n$-dimensional  
quaternion K\"ahler space $\cM^{4n}_Q$ such that its Swann bundle
coincides with $\cM^{4(n+1)}_H$. Our approach should be useful for the explicit  construction of 
new quaternion K\"ahler metrics.
The results obtained are also of  interest, e.g., in the context of 
supergravity reduction $\cN=2 \to \cN=1$, 
or alternatively  from the point of view of  embedding $\cN=1$ matter-coupled supergravity
into an $\cN=2$ theory.
\end{abstract}
\vspace{1cm}

\vfill
\end{titlepage}

\newpage
\renewcommand{\thefootnote}{\arabic{footnote}}
\setcounter{footnote}{0}

\tableofcontents{}
\vspace{1cm}
\bigskip\hrule


\section{Introduction}
\setcounter{equation}{0}

Many years ago, Bagger and Witten \cite{BW} demonstrated that 
the scalar fields of matter hypermultiplets coupled to 4D $\cN=2$ supergravity take 
their values in a $4n_{\rm H}$-dimensional
quaternion K\"ahler manifold $ \cM^{4n_{\rm H}}_Q$, 
unlike the rigid supersymmetric case where 
the  hypermultiplet target spaces are hyperk\"ahler \cite{A-GF}.
It was also pointed out in \cite{BW} that  the problem of reduction from $\cN=2$ to $\cN=1$ 
supergravity is nontrivial. For such a reduction, it is not sufficient to simply switch off 
one of the two gravitinos as well as  the graviphoton. In addition, it is  also necessary to restrict the 
scalar fields to lie in a $2n_{\rm H}$-dimensional K\"ahler-Hodge\footnote{This 
type of geometry corresponds to nonlinear couplings in $\cN=1$ supergravity \cite{WB,WB-book}.}
submanifold $ \cM^{2n_{\rm H}}_K$ 
of the $4n_{\rm H}$-dimensional quaternion K\"ahler space
 $ \cM^{4n_{\rm H}}_Q$.
Provided a  required K\"ahler-Hodge submanifold  $ \cM^{2n_{\rm H}}_K$ 
of  $ \cM^{4n_{\rm H}}_Q$ exists and 
is constructed  {\it explicitly}, the supergravity reduction  $\cN=2 \to \cN=1$ has been worked out 
by Andrianopoli, D'Auria and Ferrara \cite{ADF}, 
building in part on the mathematical results of \cite{AM}.
On the other hand, if one is interested in  embedding $\cN=1$ matter-coupled supergravity
into an $\cN=2$ theory,  one  has to ask two different 
questions that can be formulated as follows.
${}$First, given a $2n_{\rm H}$-dimensional K\"ahler-Hodge manifold $ \cM^{2n_{\rm H}}_K$, 
does there exist a quaternion K\"ahler manifold $ \cM^{4n_{\rm H}}_Q$ such 
that $ \cM^{2n_{\rm H}}_K$ is its submanifold? Second, if the answer to the first question is ``Yes,''
can one develop a regular procedure to generate $ \cM^{4n_{\rm H}}_Q$ starting 
from $ \cM^{2n_{\rm H}}_K$?
In this paper, we will argue that a natural formalism to address
these questions is the concept of rigid projective superspace \cite{KLR,LR-projective}
(see also \cite{LR2008} for a review)
and its  extension to the case of supergravity with eight supercharges elaborated in 
\cite{KTM5-Weyl,KLRT-M}.

It is known that the study of quaternion K\"ahler manifolds is 
related to that of hyperk\"ahler spaces with special properties. 
More precisely, there exists a one-to-one correspondence  \cite{Swann} (see also \cite{Galicki}) 
between $4n$-dimensional quaternion K\"ahler manifolds
and $4(n+1)$-dimensional hyperk\"ahler spaces possessing a homothetic 
Killing vector, and hence an isometric action of SU(2) rotating the complex structures. 
Such hyperk\"ahler spaces,  known in the mathematics literature as ``Swann spaces''
and often referred to as ``hyperk\"ahler cones''  in the physics literature, 
are the target spaces for rigid $\cN=2$ superconformal sigma models \cite{deWKV,deWRV}.
The above correspondence is natural from the point of view of 
the $\cN=2$ superconformal tensor calculus \cite{deWLV}, 
or more generally within the harmonic-superspace 
\cite{Galperin:1987ek,GIOS} and the projective-superspace \cite{KLRT-M}
approaches to four-dimensional $\cN=2$ matter-coupled supergravity.
In the context of $\cN=2$ supersymmetric sigma models,  
the quaternion K\"ahler manifold $ \cM^{4n_{\rm H}}_Q$ associated to 
a $4(n+1)$-dimensional hyperk\"ahler cone $ \cM^{4(n_{\rm H}+1)}_{H}$
is obtained by applying the procedure elaborated in some detail in 
\cite{deWRV} and later on applied in many publications, see, e.g., \cite{RVV,NPV}
for an incomplete list.

In the present paper, we concentrate on deriving new hyperk\"ahler cones 
with interesting geometric properties. {\it We give 
a new method for finding hyperk\"ahler cones and thus quaternion  K\"ahler manifolds,
and  also demonstrate 
 the surprising  existence of a maximal K\"ahler submanifold $ \cM^{2n_{\rm H}}_K$  
of the quaternion K\"ahler manifold $ \cM^{4n_{\rm H}}_Q$.}

In the curved projective-superspace setting, general hypermultiplet matter 
couplings to $\cN=2$ supergravity were presented in \cite{KLRT-M} and \cite{K-dual}.
The two families of locally supersymmetric sigma models 
introduced in  \cite{KLRT-M} and \cite{K-dual} are dual to each other. 
They involve the same matter hypermultiplets, which are described in terms
of $n_{\rm H} $ covariant weight-zero arctic multiplets $\U^I $ and their 
smile-conjugates\footnote{An arctic multiplet $\U$ and its smile-conjugate $\breve{\U}$
form a polar multiplet, according to the terminology introduced in  \cite{G-RRWLvU}.}  
$\breve{\U}^{\bar I}$,  
but differ in their (second) conformal compensators used.\footnote{As discussed 
in \cite{K-dual},  the sigma-model couplings  of  \cite{KLRT-M} and \cite{K-dual}
are $\cN=2$ analogues of the well-known matter couplings in the old minimal 
and the new minimal formulations for $\cN=1$ supergravity, see \cite{GGRS,BK}
for reviews.} 
In the case of the model of \cite{KLRT-M}, the compensators are 
a covariant weight-one arctic multiplet $\X$ and its smile-conjugate $\breve{\X}$. 
The compensator in the model of \cite{K-dual} is a covariant tensor multiplet $H$.
In both models, the matter $\cN=2$ superfields $\U^I $ and $\breve{\U}^{\bar I}$  take their values in 
a K\"ahler manifold $ \cM^{2n_{\rm H}}_K$  with the K\"ahler potential
$K(\Phi^{I}, {\bar \Phi}{}^{\bar{J}})$.  
Our goal in this paper is to study rigid superconformal versions 
of the models in \cite{KLRT-M,K-dual} which are obtained by retaining 
intact the compensator(s) but replacing the supergravity covariant derivatives
$\cD_A =(\cD_a, \cD_\a^i , {\bar \cD}^{\dt \a}_i)$ with those corresponding to 
a conformally flat superspace. Technically the rigid superconfomal version 
of the sigma model in  \cite{K-dual} is simpler to deal with, for the tensor compensator
$H$ is shorter\footnote{When realized in terms of $\cN=1$ superfields, 
the arctic multiplet \cite{LR-projective}
includes two physical superfields (one chiral and one complex linear)
and an infinite number of auxiliary superfields, see section 2 for more detail. 
On the contrary, the tensor multiplet consists of two physical superfields only.}
than the arctic one, $\X$. That is why we will concentrate on the study 
of this model. Some aspects of the superconformal sigma  model derived from \cite{KLRT-M}
were studied in \cite{K-hyper} where that model was first introduced.

This paper is organized as follows. In section 2 we introduce the off-shell 
$\cN=2$ superconformal sigma model to be studied and discuss its geometric aspects.
The main thrust  of section 3 is to argue that the off-shell  $\cN=2$ superconformal symmetry
of the model can be used to convert the infinite set of algebraic auxiliary field equations into 
a single second-order differential equation under given initial conditions, which 
is a deformation of the geodesic equation, with the complex coordinate for ${\mathbb C}P^1$
being the evolution parameter. 
In section 4 we explicitly eliminate the auxiliary superfields and derive the
hypermultilet Lagrangian in terms of the physical superfields, 
in the case when $ \cM^{2n_{\rm H}}_K$ is chosen to be ${\mathbb C}P^{n_{\rm H}}$.
Section 5 is devoted to the discussion of the results obtained. 
Two technical appendices are also included. 
In Appendix A we  list the $\cN=2$ superconformal transformations of several off-shell 
supermultiplets and their relization in $\cN=1$ superspace.  
Appendix B  contains a few results concerning the $\cN=2$ supersymmetric 
sigma models on (co)tangent bundles of Hermitian symmetric spaces.

\section{The sigma model and its geometric properties}
\setcounter{equation}{0}
In this paper we are interested in a rigid superconformal version 
of the four-dimensional  $\cN=2$ locally supersymmetric model proposed in \cite{K-dual}.
This theory is formulated in $\cN=2$ projective superspace \cite{LR-projective}, 
and therefore its action can naturally be written either in terms of $\cN=2$ 
projective superfields or in terms of the associated $\cN=1$ superfields.
We will use both realizations in different parts of this paper, 
and the latter will be used to formulate the action.
It consists of two terms, 
\be
S [H(\z), \U (\z)] = \k S_{\rm T} +  S_{\rm H}, 
\label{nact}
\ee
where $\k$   is a  constant parameter, and 
\bea
S_{\rm T} &=& - 
\oint \frac{{\rm d}\z}{2\pi {\rm i} \z} \,  
 \int  {\rm d}^4 x \,{\rm d}^4\q\,
 H \ln H ~,
 \label{act-tensor} \\
 S_{\rm H} &=&  
  \oint \frac{{\rm d}\z}{2\pi {\rm i} \z} \,  
 \int  {\rm d}^4 x \, {\rm d}^4\q\,H \,  K \big( \U^I , \breve{\U}^{\bar J}   \big) ~,
\label{act-hyper} 
\eea
with some closed integration contours in the $\z$-plane.
Here $H(\z) $ is an $\cO(2)$ multiplet \cite{KLR} 
(or $\cN=2$ tensor multiplet \cite{Wess}) 
\bea
H(\z) = \frac{1}{\z}\, \vf + G - \z \,{\bar \vf}~, \qquad {\bar D}_{\dt \a} \vf =0~, 
\qquad {\bar D}^2 G =0~, \quad {\bar G}=G~, 
\label{tensor-series}
\eea
$\U^I (\z)$  a set of arctic hypermultiplets \cite{LR-projective}, $I=1, \dots, n_{\rm H}$, 
\be
 \U^I (\z) = \sum_{n=0}^{\infty}  \, \z^n \U_n^I  = 
\F^I + \z \, \S^I  + O(\z^2) ~,\qquad
{\bar D}_{\dt{\a}} \F^I =0~, \quad {\bar D}^2 \S = 0 ~,
\ee
and 
$\breve{\U}^{\bar I} (\z)$ their smile-conjugates 
\be
\breve{\U}^{\bar I} (\z) = \sum_{n=0}^{\infty}  \,  (-\z)^{-n}\,
{\bar \U}_n^{\bar I}~.
\ee
The $\cN=1$ superfields $\U^I_2$, $\U^I_3, \dots $, are complex unconstrained.
Since these appear in the action without derivatives, they are purely auxiliary degrees of freedom. 
The  hypermultiplet action $ S_{\rm H} $ involves the  K\"ahler potential, 
$K(\Phi^{I}, {\bar \Phi}{}^{\bar{J}})$, 
of a real-analytic K\"ahler manifold $\cM \equiv \cM^{2n_{\rm H}}_K$ 
of complex dimension $n_{\rm H}$.

The action $S_{\rm T}$
is the $\cN=2$ projective-superspace formulation \cite{KLR}
of the $\cN=2$ improved  tensor multiplet model \cite{deWPV}.
Its realization in terms of $\cN=1$ 
superfields  was first developed in  \cite{LR}:
\bea
S_{\rm T}  
  =    \int  {\rm d}^4 x \,{\rm d}^4\q 
  \,L_{\rm T}(G, \vf, \bar \vf ) ~, \qquad 
  L_{\rm T}(G, \vf, \bar \vf) =
    {\mathbb H} - G \, \ln \big( G+  {\mathbb H}\big)   ~, 
\label{Lag-tensor}
\eea
where 
\bea
{\mathbb H}:= \sqrt{ G^2 +4\vf \bar \vf }~.
\eea
The combination $G+  {\mathbb H}$ 
 naturally originates, e.g.,  as follows:
\bea
H(\z) = \hf (G+ {\mathbb H}) \Big( 1- \z \, \frac{2\bar \vf}{G+{\mathbb H}} \Big)
 \Big( 1+ \frac{1}{\z} \, \frac{2\vf}{G+{\mathbb H}} \Big)~.
\label{identity} 
\eea

The theory with action (\ref{nact}) is $\cN=2$ superconformal provided
$H(\z)$ transforms as a $\cN=2$ superconformal tensor multiplet and 
$\U^I$ as a superconformal  weight-zero arctic multiplet \cite{K-hyper}. 
The corresponding transformation laws 
are given below in  eqs.  (\ref{sctl-H}) and (\ref{sctl-U}) respectively.

As discussed in \cite{K-dual}, the theory (\ref{nact}) possesses a 
dual formulation obtained by dualizing the tensor multiplet $H(\z)$
into an arctic multiplet $\X (\z)$ and its conjugate 
following the procedure given, e.g., in  \cite{G-RRWLvU}. 
The resulting hypermultiplet sigma model \cite{K-hyper}
\bea
 S_{\rm dual} [ \X (\z) , \U (\z)]&=&  
  \k\, \oint \frac{{\rm d}\z}{2\pi {\rm i} \z} \,  
 \int  {\rm d}^4 x \, {\rm d}^4\q\, \breve{\X} \,\X \,
 \exp \Big\{  \frac{1}{\k} \,K ( \U , \breve{\U}   ) \Big\} ~
\label{act-dual} 
\eea
is $\cN=2$ superconformal provided 
$\X$ transforms  as the superconformal weight-one arctic multiplet, 
see Appendix \ref{App1} for the corresponding transformation law.
The above theory is the rigid superconformal  limit of the locally supersymmetric 
sigma model proposed in \cite{KLRT-M}. On the other hand, the theory with action 
(\ref{nact}) is the rigid superconformal  version of the locally supersymmetric 
sigma model proposed in \cite{K-dual}.

As pointed out  in  \cite{K-dual}, the theory (\ref{nact}) is a natural extension 
of the $\cN=1$ superconformal sigma model:
 \bea
 S[G, \F ] = -\k  \int  {\rm d}^4 x \,{\rm d}^4\q\,
 G   \ln G   +   \int  {\rm d}^4 x \,{\rm d}^4\q\,
 G\,  K(\Phi^{I}, {\bar \Phi}{}^{\bar{J}})~.
\label{N=1act}
\eea
Here the first term is proportional to the action for the $\cN=1$ improved 
tensor multiplet \cite{deWR}. 
The dual version of (\ref{N=1act}) is 
\bea
 S_{\rm dual} [ \c , \F] = k  \int  {\rm d}^4 x \,{\rm d}^4\q\, {\bar \c}\c\,
  \exp \Big\{  \frac{1}{\k} \,  K(\Phi, {\bar \Phi}) \Big\}~,
\label{N=1act-dual}
\eea
with $\c$ a chiral scalar superfield.
As is known, the action $ S_{\rm dual} [ \c , \F] $  is obtained from 
that describing chiral matter in $\cN=1$ supergravity
(see, e.g., \cite{GGRS,BK} for  reviews)
by switching off the (axial) vector gravitational superfield and
keeping intact the chiral compensator $\c$ and its conjugate.
Clearly, the superconformal  sigma model (\ref{act-dual}) is 
an $\cN=2$ extension of (\ref{N=1act-dual}). 

The extended superconformal  sigma model  (\ref{nact}) 
inherits  all the geometric features of
its $\cN=1$ predecessor (\ref{N=1act}). 
The K\"ahler invariance of the latter,
\be
K(\F, \bar \F) \quad \longrightarrow \quad K(\F, \bar \F)~ +~
F(\F)+  {\bar F} (\bar \F) 
\label{kahl}
\ee
turns into 
\be
K(\U, \breve{\U})  \quad \longrightarrow \quad K(\U, \breve{\U}) ~+~
F(\U) \,+\, {\bar F} (\breve{\U} ) 
\label{kahl2}
\ee
for the model (\ref{nact}), where we have used the identity
\bea
 \oint \frac{{\rm d}\z}{\z} \,  
 \int  {\rm d}^4 x \, {\rm d}^4\q\,H \,  F (\U)=0~, 
 \eea
 for any holomorphic function $F (\F)$.
A holomorphic reparametrization of the K\"ahler manifold $\cM$,
\be
 \F^I  \quad  \longrightarrow   \quad f^I \big( \F \big) ~,
\ee
has the following counterpart
\be
\U^I (\z) \quad  \longrightarrow  \quad f^I \big (\U(\z) \big)
\label{kahl3}
\ee
in the $\cN=2$ case. Therefore, the physical $\cN=1$
superfields of the $\cN=2$ arctic multiplet
\be
 \U^I (\z)\Big|_{\z=0} ~=~ \F^I ~,\qquad  \quad \frac{ {\rm d} \U^I (\z) 
}{ {\rm d} \z} \Big|_{\z=0} ~=~ \S^I ~,
\label{kahl4} 
\ee
should be regarded, respectively, as  coordinates of a point in the K\" ahler
manifold and a tangent vector at  the same point. 
Thus the variables $(\F^I, \S^J)$ parametrize the holomorphic tangent 
bundle $T\cM$ of the K\"ahler manifold $\cM$. 
This interpretation of the physical variables of the hypermultiplet theory (\ref{act-hyper})
coincides with that proposed in  \cite{K-double} for the 
{\it non-superconformal} sigma model
\bea
  S [\U(\z)]&=& 
  \oint \frac{{\rm d}\z}{2\pi {\rm i} \z} \,  
 \int  {\rm d}^4 x \, {\rm d}^4\q\,  K \big( \U^I , \breve{\U}^{\bar J}   \big) ~.
\label{H=1} 
\eea
which is obtained from  (\ref{nact}) by ``freezing'' the tensor multiplet, 
that is by replacing $H(\z)$ with its $\z$-independent 
expectation value $\langle H \rangle={\rm const}$.  

Suppose that in the action (\ref{act-hyper}) 
we have eliminated all the auxiliary superfields contained in $\U$ and $\breve \U$
with the aid of the corresponding algebraic equations of motion
\bea
\oint \frac{{\rm d} \z}{\z} \,\z^n \, H \,
\frac{\pa K(\U, \breve{\U} ) }{\pa \U^I} 
~ = ~ \oint \frac{{\rm d} \z}{\z} \,\z^{-n} \, H\, \frac{\pa 
K(\U, \breve{\U} ) } {\pa \breve{\U}^{\bar I} } ~ = ~
0 ~. \qquad n \geq 2 ~                
\label{asfem}
\eea
Let $\U_*(\z) \equiv \U_*( \z; \F, {\bar \F}, \S, \bar \S )$ 
denote their unique solution subject to the initial conditions (\ref{kahl4}) 
\bea
\U_* (0)  = \F ~,\qquad  \quad \dt{\U}_* (0) 
 = \S ~.
\label{geo3} 
\eea
The action (\ref{nact}) then turns into
\bea
S [G, \vf , \F , \S] &:=&  S [H (\z) , \U_* (\z) ] 
= \int  {\rm d}^4 x \,{\rm d}^4\q 
  \,L(G, \vf, {\bar \vf}, \F , {\bar \F}, \S, {\bar \S})~, \non \\
L(G, \vf, {\bar \vf}, \F , {\bar \F}, \S, {\bar \S}) &=&
\k\, L_{\rm T}(G, \vf, \bar \vf) 
+L_{\rm H}(G, \vf, {\bar \vf}, \F , {\bar \F}, \S, {\bar \S})~ . 
\label{nact_2}
\eea
Here the tensor multiplet Lagrangian is given by eq. (\ref{Lag-tensor}).
In accordance with the generalized Legendre transform procedure \cite{LR-projective},
we should  dualize the real linear superfield $G$ into
a chiral scalar $\c$ and its conjugate $\bar \c$, and further dualize 
the complex linear tangent variables $\S^I$ and their conjugates
${\bar \S}^{\bar I}$ into chiral one-forms $\J_I$ and their conjugates 
${\bar \J}_{\bar I}$, ${\bar D}_{\dt \a} \c = {\bar D}_{\dt \a} \J_I =0$.
This results in 
\bea
S [G, \vf , \F , \S] \quad \longrightarrow \quad S [\c, \vf , \F , \J]  =  
\int  {\rm d}^4 x \, {\rm d}^4\q\,  \cH  \big(\c, {\bar \c}, \vf , {\bar \vf}, 
\F^I , {\bar \F}^{\bar J}, \J_I , {\bar \J}_{\bar J}\big)~.~~~
\label{nact_3}
\eea  
{\em We thus have the following striking situation}:
The target space of this sigma model is a hyperk\"ahler manifold, 
$HKC^{4(n_{\rm H} +1)} (\cM)$,
of real dimension $4n_{\rm H} +4$. 
Since the sigma model is $\cN=2$ superconformal,
$HKC^{4(n_{\rm H} +1)} (\cM)$ is a hyperk\"ahler cone
in the sense of \cite{deWKV,deWRV}.
The Lagrangian $\cH$ in (\ref{nact_3})
is the hyperk\"ahler potential for  $HKC^{4(n_{\rm H} +1)} (\cM)$.
Note that the variables $(\F^I, \J_J)$ 
parametrize the holomorphic cotangent 
bundle $T^*\cM$ of the K\"ahler manifold $\cM= \cM^{2n_{\rm H} }_K$. 
As mentioned in the introduction, there exists a one-to-one
correspondence between $4n$-dimensional quaternion K\"ahler spaces (QK)
and $4(n+1)$-dimensional hyperk\"ahler cones  (HKC) \cite{Swann,Galicki}.
In our case $HKC^{4(n_{\rm H} +1)} (\cM) \longleftrightarrow  QK^{4n_{\rm H} } (\cM)$. 
The K\"ahler manifold $\cM^{2n_{\rm H} }_K$ is embedded into $QK^{4n_{\rm H} } (\cM)$. \\

\section{Superconformal invariance and the auxiliary field equations} 
\setcounter{equation}{0}

When dealing with the $\cN=2$ off-shell superconformal sigma-model (\ref{nact}), 
the main technical challenge is to explicitly eliminate the auxiliary 
superfields $\U^I_2$, $\U^I_3, \dots $, 
by means of solving the corresponding equations of motion (\ref{asfem}).
This section is devoted to a general analysis of the problem.

\subsection{Superconformal invariance}

Both  actions (\ref{act-tensor}) and (\ref{act-hyper}) are $\cN=2$ superconformal. 
To write down the superconformal transformations 
of $H(\z)$, $\U (\z) $ and $\breve{\U}(\z)$, it is useful to lift these multiplets
to $\cN=2$ superspace ${\mathbb R}^{4|8}$ parametrized  
by  coordinates  $ z^A = (x^a,  \q^\a_i, {\bar \q}^i_\dt{\a} )$, 
where $i=\1, \2 $. 
In the $\cN=2$ setting, each of  $H(\z)$, $\U (\z) $ and $\breve{\U}(\z)$
is a projective multiplet. In general, 
with respect to the $\cN=2$ Poincar\'e supersymmetry, 
a projective multiplet 
$Q(\z)$ is determined by the two conditions \cite{LR-projective}:\\
(i) it is characterized by two fixed integers $p,q$ (of which $p$ may be  equal to $- \infty$
and $q$ to  $+\infty$) such that
\be
Q(z, \z) = \sum_{p}^{q} Q_n(z) \z^n~;
\label{series}
\ee
(ii)  it is subject to the constraints
\bea
D_\a (\z) Q(\z)= {\bar D}_{\dt \a} (\z)Q(\z) =0~, 
\label{constraints}
\eea
where
\be
D_\a (\z):= \z^i D_{\a i} ~, \qquad  {\bar D}_{\dt \a} :=  \z^i {\bar D}_{{\dt \a}i} ~, 
\qquad \z^i:=(1,\z)~,
\ee 
where $D_A = (\pa_a, D^i_\a, {\bar D}^{\dt \a}_i)$ are the $\cN=2$ 
flat covariant derivatives.
With respect to the $\cN=2$ superconformal group, 
the admissible transformation laws prove to depend on the parameters 
$p$ and $q$ in (\ref{series}) as shown in  \cite{K-hyper}.

The following remark is needed here.
It follows from the constraints (\ref{constraints}) 
that the dependence of $Q(x, \q_i, {\bar \q}^i , \z)$ 
on the Grassmann variables $\q^\a_{\2}$ and ${\bar \q}^{\2}_{\dt \a}$ 
is uniquely determined in terms 
of its dependence on $\q^\a_{\1}\equiv \q^\a$
and ${\bar \q}^{\1}_{\dt \a}\equiv {\bar \q}_{\dt \a}$.  In other words, 
the projective superfields depend effectively 
on half the Grassmann variables which can be chosen
to be the spinor coordinates of  $\cN=1$ superspace.
In other words, no information is lost if we replace $Q(\z)$ 
by its $\cN=1$ projection $Q(\z)|$ defined as 
\be
U| = U(x,\q_i, {\bar \q}^i)\Big|_{\q_{\2}={\bar \q}^{\2}=0}~,
\ee
for any $\cN=2$ superfield $U(x,\q_i, {\bar \q}^i)$.

The actions (\ref{act-tensor}) and (\ref{act-hyper}) are 
invariant under the following 
$\cN=2$ superconformal transformations 
of $H$, $\U$ and $\breve \U$ 
\cite{K-hyper}: 
\begin{subequations}
\bea
\z\,  \d  H &=&  
- \Big(  \x +  {\l}^{++} (\z) \,\pa_\z \Big) (\z H )
-2 \,\S (\z)\, \z H ~, 
\label{sctl-H} \\
\d \U^I &=& - \Big(  \x +  {\l}^{++} (\z)\,\pa_\z \Big) \U^I ~, \qquad
\d \breve{\U}^{\bar I} = - \Big(  \x +  {\l}^{++} (\z)\,\pa_\z \Big) \breve{\U}^{\bar I} ~.
\label{sctl-U}
\eea
\end{subequations}
Here $\x$ is a $\cN=2$ superconformal Killing vector, 
\be
\x = {\overline \x} =\x^A (z)D_A
= \x^a (z) \,\pa_a + \x^\a_i (z)\,D^i_\a
+ {\bar \x}_{\dt \a}^i (z)\, {\bar D}^{\dt \a}_i~,
\ee   
with the master property 
\bea
{\bar D}^{\dt \a}_i \J =0 \quad \longrightarrow \quad 
{\bar D}^{\dt \a}_i (\x \,\J )=0~, 
\eea
for any chiral superfield $\J$.
The superconformal parameters $ {\l}^{++} (\z)$ and $\S (\z)$ 
appearing in (\ref{sctl-H}) and (\ref{sctl-U})  
have the form 
\bea
{\l}^{++} (\z)&=& 
=  {\l}^{\1 \1 }\, \z^2 -2  {\l}^{\1 \2}\, \z + {\l}^{\2 \2} ~,
\qquad \S(\z)=  - {\l}^{\1 \1} \,\z + {\l}^{\1 \2} + {\s} +\bar \s ~
\eea
in terms of the descendants $\s$ and $\l^{ij}$  of $\x$  defined as
\bea
\s &=& \frac{1}{4}
{\bar D}^{\dt \a}_i {\bar \x}_{\dt \a}^{ i} ~, 
\qquad {\bar D}^{\dt \a}_{ i} {\s} ~=~0
\non \\
{\l}_j{}^i  &=& \hf \Big(D^i_\a \x^\a_j - \hf \d^i_j D^k_\a \x^\a_k \Big)
~, \qquad 
\l^{ij}=\l^{ji}~,
\quad \overline{\l^{ij} } = \l_{ij}  ~.
\eea
It should be remarked that these descendants originate as follows 
\be
[\x \;,\; D^i_\a ] = - (D^i_\a \x^\b_j) D^j_\b
= { \o}_\a{}^\b  D^i_\b - 
\bar{ {\s}} \, D^i_\a
- {\l}_j{}^i \; D^j_\a \quad \longrightarrow \quad
{\bar D}^{\dt \a}_i \x^\b_j =0~,
\label{4DmasterN=2} 
\ee
where 
\be
{\o}_{\a \b} = -\frac{1}{2}\;D^i_{(\a} \x_{\b)i}~, \qquad
{\bar D}^{\dt \a}_{ i} {\o}_{\a \b}= 0~. 
\ee
See  Refs. \cite{HH,Park,KT,K-hyper} for more detail about superconformal 
transformations in $\cN=2$ superspace.

The superconformal transformation of $H(\z)$, eq. (\ref{sctl-H}),
 proves to be  uniquely determined
by the constraints obeyed by this multiplet, $D_\a (\z) H(\z)= {\bar D}_{\dt \a} (\z)H(\z) =0,$
and by its explicit dependence of $\z$ given by (\ref{tensor-series}). 
Eq. (\ref{sctl-U}) means that $\U^I(\z)$ is a weight-zero 
arctic multiplet. 
The superconformal transformations of the weight-$n$ arctic and
antarctic multiplets are given by eqs. (\ref{arctic2}) and
(\ref{antarctic2}) respectively. 
 
Consider the  model (\ref{act-dual}) dual to (\ref{nact}). 
As discussed in section 2, it is $\cN=2$ superconformal invariance 
provided $\U^I$ and 
$\X$ transform as the weight-zero and  weight-one arctic multiplets, 
respectively, with the latter transformation law given by eq. (\ref{arctic2}) with $n=1$.
 
It is of interest to analyze the superconformal properties of the auxiliary field equations (\ref{asfem}). 
In complete analogy with the case $H=1$ \cite{GK,LR2008}, 
these equations imply that 
\bea
\O_I  (\z) := \z \, H\, \frac{\pa K(\U, \breve{\U} ) }{\pa \U^I} 
\label{Omega}
\eea
has no poles in $\z$ and  therefore can be represented  by a Taylor series 
\bea
\O_I  (\z)= \sum_{n=0}^{\infty}  \, \z^n \O_{n\,I} ~.
\label{asfem2}
\eea
This superfield becomes an arctic multiplet on the full mass shell 
when  the equations of motion for $\F^I$ and $\S^I$ are  imposed as well.

Let us promote the superfields $H$, $\U$ and $\breve \U$ in (\ref{Omega}) 
to  $\cN=2$ projective superfields.
Then, using the transformation laws (\ref{sctl-H}) and (\ref{sctl-U}), we observe that
the  composite (\ref{Omega}) transforms as 
\bea
\d \O_I = - \Big(  \x &+&  {\l}^{++} (\z)\,\pa_\z \Big) \O_I
- 2\,\S (\z) \, \O_I~.
\label{sctl-O}
\eea
It is a simple exercise to check that this transformation law preserves 
the functional form of $\O_I$  given in  (\ref{asfem2}).
Therefore, the auxiliary field equations (\ref{asfem}), or equivalently (\ref{asfem2}),
are   $\cN=2$ superconformal. 
On the full mass shell, eq. (\ref{sctl-O}) tells us that $\O_I$ is a weight-two 
superconformal arctic multiplet. 
 
We wish to convert the algebraic auxiliary field equations (\ref{asfem})
into an equivalent  second-order ordinary differential equation obeyed by $\U(\z)$, 
with $\z$ the evolution parameter. This is certainly possible in the case $H=1$, 
as has been shown in \cite{GK,GK2} for the Hermitian symmetric spaces, and 
in \cite{K-hyper} for general K\"ahler spaces. In the case of an arbitrary 
$\cN=2$ tensor multiplet $H(\z)$,  let us proceed to derive 
such an equation for a simplest K\"ahler potential.

\subsection{Quadratic K\"ahler potential}
\label{quadratic}

The auxiliary  field equations  (\ref{asfem}) 
can be explicitly solved in the case 
of a flat K\"ahler  target space described by the potential\footnote{Our consideration can 
be trivially generalized to the case of an indefinite metric in (\ref{quadpot}), $\d_{I{\bar J}} \to \eta_{I{\bar J}}$.}
\be
{\mathfrak K}(\F, \bar \F ) = \F^\dagger \F = \d_{I{\bar J}} \,\F^I{\bar \F}^{\bar J} ~.
\label{quadpot}
\ee
Then, using eq. (\ref{identity})  we find
\bea
H \,{\mathfrak K}(\U, \breve{ \U })  &=& H(\z) \breve{ \U}^{\bar I}(\z)  \U^I (\z) 
= \hf (G+ {\mathbb H})  \breve{\bm  \U}{}^{\bar I} (\z) {\bm \U}^I (\z)~,  
\eea
where
\bea
 {\bm \U}^I (\z)&:=&  \Big( 1- \z \, \frac{2\bar \vf}{G+{\mathbb H}} \Big)\U^I (\z) 
 =\sum_{n=0}^{\infty} \z^n  {\bm \U}^I_n ~.
\eea
The new superfields, $ {\bm \U}^I (\z)$, are not arctic, for the components 
\be
 {\bm \U}^I_1 = \S^I - \frac{2\bar \vf}{G+{\mathbb H}}  \,\F^I
 \ee
 obey a modified linear constraint. The new auxiliary superfields 
 $ {\bm \U}^I_2,   {\bm \U}^I_3, \dots$, can be immediately eliminated. 
 As a result,  the Lagrangian corresponding to the hypermultiplet action  (\ref{act-hyper}) 
with $K = \mathfrak K$  becomes
 \bea
 {\mathfrak L}_{\rm H} &=& \frac{1}{2} (G+ {\mathbb H}) \Big\{ {\bar \F}^{\bar I} \F^I
 -  \big( {\bar \S}^{\bar I} - \frac{2 \vf }{ G+{\mathbb H} }  \,{\bar \F}^{\bar I} \Big) 
  \Big( \S^I - \frac{2\bar \vf}{G+{\mathbb H}}  \,\F^I \Big) \Big\}~.
 \eea
In terms of the K\"ahler potential  ${\mathfrak K}(\F, \bar \F ) $, 
this Lagrangian can equivalently be rewritten in the form:
\bea
{\mathfrak L}_{\rm H}= G\, {\mathfrak K} + \varphi {\mathfrak K}_I\Sigma^I 
+\bar\varphi {\mathfrak K}_{\bar J}\bar \Sigma^{\bar J}
-\frac12 (G+ {\mathbb H}) {\mathfrak K}_{I\bar J}\Sigma^I\bar\Sigma^{\bar J}~.
\eea

Let $\U_*^I(\z)$ denote the unique solution to the {\it algebraic} auxiliary field equations
under the initial conditions (\ref{geo3}). It has the form
\be
\U_*^I(\z) = \F^I + \frac{\z}{ 1- {\bar \L} \z}\, \S^I~,
\ee
where 
\be
 { \L} := \frac{2 \vf}{G+{\mathbb H}} ~.
\ee
It is further a solution to  the following {\it differential} equation:
\bea
 \frac{ {\rm d}^2 \U^I (\z) }{ {\rm d} \z^2 } 
- 2\, \frac{\bar \L}{ 1- {\bar \L} \z}\,\frac{ {\rm d} \U^I (\z) }{ {\rm d} \z } =0 ~.
\label{geodesic3}
\eea

It is instructive to check that equation (\ref{geodesic3}) is superconformal. 
Introduce the following superfield:
\bea
\P^I := \frac{ {\rm d}^2 \U^I (\z) }{ {\rm d} \z^2 } 
- 2\, \frac{\bar \L}{ 1- {\bar \L} \z}\,\frac{ {\rm d} \U^I (\z) }{ {\rm d} \z } ~.
\label{X}
\eea
We are going to demonstrate that its superconformal transformation is
\bea
\d \P^I &=& - \Big(  \x +  {\l}^{++} \,\pa_\z + 2\big(\pa_\z  {\l}^{++}\big)
\Big) \P^I ~.
\label{sctl-X}
\eea
Using eq. (\ref{sctl-U}) gives
\begin{subequations}
\bea
 \frac{ {\rm d} }{ {\rm d} \z } \d \U^I 
& =&  - \Big(  \x +  {\l}^{++} \,\pa_\z + \big(\pa_\z  {\l}^{++}\big)
\Big)  \frac{ {\rm d} }{ {\rm d} \z } \d \U^I  ~,  \\
\frac{ {\rm d}^2 }{ {\rm d} \z^2 } \d \U^I 
& =&  - \Big(  \x +  {\l}^{++} \,\pa_\z + 2 \big(\pa_\z  {\l}^{++}\big)
\Big)  \frac{ {\rm d^2} }{ {\rm d} \z^2 } \d \U^I  
-  \big(\pa^2_\z  {\l}^{++}\big)  \frac{ {\rm d} }{ {\rm d} \z } \d \U^I  ~.
\eea
\end{subequations}
Next, making use of eq. (\ref{sctl-H}) allows us to read off the superconformal transformations 
of the components of $H(\z)$: 
\begin{subequations}
\bea
\d G &=& -\x G -2 (\s +{\bar \s} )G  + 2\l^{\2\2}{\bar \vf} +2 \l^{\1\1} \vf ~, 
\label{sctl-G}\\
\d \vf  &=& -\x \vf -2 (\s +{\bar \s} )\vf  - \l^{\2\2} G -2 \l^{\1\2} \vf ~, 
\label{sctl-vf} \\
\d {\bar \vf}   &=& -\x {\bar \vf} -2 (\s +{\bar \s} ){\bar \vf}  
- \l^{\1\1} G +2 \l^{\1\2} \vf ~.
\label{sctl-bar-vf}
\eea
\end{subequations}
These results immediately lead to 
\bea
\d {\mathbb H} = -\x  {\mathbb H} -2 (\s +{\bar \s} ) {\mathbb H} ~,
\eea
as well as to
\bea
\d { \L} = -\x \L
-\l^{\2\2} - 2\l^{\1\2} { \L} -\l^{\1\1} { \L}^2 ~, \qquad
\d {\bar \L} = -\x {\bar \L} -\l^{\1\1} + 2\l^{\1\2} {\bar \L} -\l^{\2\2} ({\bar \L})^2 ~.~~~
\eea
Making use of the results obtained, we check  that 
\bea
 \big(\pa^2_\z  {\l}^{++}\big)
+ \Big\{ \l^{++} \pa_\z  
+\big(\pa_\z  {\l}^{++}\big)\Big\}  \frac{2 \bar \L}{ 1- {\bar \L} \z} 
+   \frac{2 (\d {\bar \L} +\x{\bar \L})}{ (1- {\bar \L} \z)^2}  =0~.
\eea
The above identities indeed justify 
the superconformal transformation law (\ref{sctl-X}), 
and hence the fact that the differential equation (\ref{geodesic3}) is superconformal. 

It should be pointed out that the hypermultiplet model (\ref{act-hyper}) 
with K\"ahler potential (\ref{quadpot}) possesses 
a dual off-shell formulation obtained by dualizing each\footnote{The other option is
to dualize only a subset of the $n_{\rm H}$ polar multiplets.}  
polar multiplet, 
 $\U^I$ and $\breve{\U}^{\bar I}$, into a real $\cO (2)$  multiplet
$\eta^I$, with $I =1, \dots, n_{\rm H}$.\footnote{As is well-known \cite{LR}, this is possible only if 
the model possesses an isometry so that it does not depend on the phase of $\Upsilon$.} 
The dual formulation  is described by  the following $\cN=2$ superconformal action: 
\bea
{\mathfrak S}_{\rm H, dual} = - 
\oint \frac{{\rm d}\z}{2\pi {\rm i} \z} \,  
 \int  {\rm d}^4 x \, {\rm d}^4\q\,
\frac{\eta^I \eta^I}{2 H} ~.
\label{universal}
\eea
Of the global  $U(n_{\rm H}) $ symmetry of the original hypermultiplet action, 
only its subgroup $O(n_{\rm H}) $  is manifestly realized  in the dual formulation, 
while the other symmetries emerge as  duality transformations. 
In the same vein, of the $2n_{\rm H}$ Abelian symmetries 
\bea
\d H (\z) =0 ~, \qquad \d \U^I (\z) = c^I ={\rm const}
\eea
of the original hypermultiplet model, only $n_{\rm H}$ 
(Peccei-Quinn-type) symmetries are manifestly realized in 
the dual formulation:
 \bea
\d H (\z) =0 ~, \qquad \d \eta^I (\z) = H(\z) \,a^I ~, \qquad 
a^I =\overline{a^I} = {\rm const}~.
\eea
Modulo sign, the sigma model (\ref{universal}) with $n_{\rm H} =1$ 
is known to define the hyperk\"ahler cone 
corresponding to the classical universal hypermultiplet \cite{BS,deWRV}.

\subsection{Modified geodesic equation}

If the K\"ahler space is not flat, the differential equation (\ref{geodesic3})
is no-longer equivalent to the auxiliary field equations  (\ref{asfem}).
Guided by the experience gained in the case $H=1$ \cite{GK,GK2,AKL1,K-hyper}
we should look for a generalization of 
eq. (\ref{geodesic3})  of  the form:
\bea 
{\bm \P}^I = 0~,
\label{geodesic}
\eea
where
\bea
{\bm \P}^I &:=&
\P^I + \G^I_{~JK} \Big( \U (\z), \bar{\F} \Big)\,
\frac{ {\rm d} \U^J (\z) }{ {\rm d} \z } \,
\frac{ {\rm d} \U^K (\z) }{ {\rm d} \z }  +\dots \non \\
&\equiv&  \frac{ {\rm d}^2 \U^I (\z) }{ {\rm d} \z^2 }-  \frac{2 \bar \L}{ 1- {\bar \L} \z}\,
\frac{ {\rm d} \U^I (\z) }{ {\rm d} \z }  
+ 
\G^I_{~JK} \Big( \U (\z), \bar{\F} \Big)\,
\frac{ {\rm d} \U^J (\z) }{ {\rm d} \z } \,
\frac{ {\rm d} \U^K (\z) }{ {\rm d} \z }  +
\D {\bm \P}^I~.~~~~
\label{J} 
\eea
Here the term containing the Christoffel symbol $  \G^I_{~JK} $ 
is required to ensure the correct transformation of ${\bm \P}^I $
under holomorphic reparametrizations (\ref{kahl3}).
It can be argued that the last term in (\ref{J})  must depend on the K\"ahler potential only
via the corresponding K\"ahler metric, 
the Riemann tensor and its covariant derivatives.  
The superfield ${\bm \P}^I$ should be chosen 
such that\\
${}\quad$ (i) in the case when $H=1$ and  the K\"ahler manifold $\cM^{2n_{\rm H}}_K$ is 
Hermitian symmetric,  \\
\phantom{${}\quad$ (i)} eq. (\ref{geodesic}) should 
reduce to the  geodesic equation (\ref{OldGeodesic});\\
${}\quad$ (ii) the  $\cN = 2$ superconformal transformation of ${\bm \P}^I$ should be
\bea
\d {\bm \P}^I &=& - \Big(  \x +  {\l}^{++} \,\pa_\z + 2\big(\pa_\z  {\l}^{++}\big)
\Big) {\bm \P}^I ~.
\label{sctl-J}
\eea

It turns out that the above requirements allow one, in principle,  to reconstruct
$\D {\bm \P}^I$ in (\ref{J}) step by step in perturbation theory. 
As a first step, varying the right-hand side of (\ref{J}) gives
\bea
\d {\bm \P}^I &=& - \Big(  \x +  {\l}^{++} \,\pa_\z + 2\big(\pa_\z  {\l}^{++}\big)
\Big) {\bm \P}^I  \non \\
&&-\l^{\1\1} \,
R_{J {\bar L}K}{}^I \Big( \U (\z), \bar{\F} \Big)\,
{\bar \S}^{\bar L} \,\frac{ {\rm d} \U^J (\z) }{ {\rm d} \z } \,
\frac{ {\rm d} \U^K (\z) }{ {\rm d} \z }  +\dots
\label{sctl-X2}
\eea
To derive this  and some other relations below, one 
has to use the superconformal transformations of $\F$, $\S$ and their conjugates:
 \begin{subequations} 
 \bea
 \d \F^I  &=& -\x \F^I - \l^{\2\2} \S^I ~, 
 \label{sctl-F} \\
 \d {\bar \F}^{\bar I}  &=& -\x {\bar \F}^{\bar I}  - \l^{\1\1} {\bar \S}^{\bar I}  ~,
  \label{sctl-bar-F}  \\
\d \S^I  &=& -\x \S^I +2\l^{\1\2} \S^I   - \l^{\2\2} \U_2^I ~,
 \label{sctl-S}  \\
\d {\bar \S}^{\bar I}  &=& -\x {\bar \S}^{\bar I}  -2\l^{\1\2} {\bar \S}^{\bar I} 
- \l^{\1\1} {\bar \U}_2^{\bar I}  ~.
 \label{sctl-bar-S} 
 \eea
 \end{subequations}
 These relations follow from (\ref{sctl-U}). In (\ref{sctl-S})  and (\ref{sctl-bar-S}), 
 $\U_2^I$ and its conjugate should be expressed in terms of the physical 
 superfields $\F^I$, $\S^I$ and their conjugates,  in accordance with
 (\ref{geodesic}).
 
To cancel the variation in the second line of (\ref{sctl-X2}), it can be shown
that  the last term in (\ref{J}) should  have the form:
 \bea
 \D {\bm \P}^I = -\frac{\bar \L}{1+ \L \bar \L } \,
R_{J {\bar L}K}{}^I\Big( \U (\z), \bar{\F} \Big)\,
{\bar \S}^{\bar L} \,\frac{ {\rm d} \U^J (\z) }{ {\rm d} \z } \,
\frac{ {\rm d} \U^K (\z) }{ {\rm d} \z }  + O(R^2, \nabla R) ~.
\label{cubic}
\eea
Here $ O(R^2, \nabla R)$ denotes terms of second and higher orders in the target space curvature, 
or terms containing covariant derivatives of  the target space curvature.  

\subsection{Leading contributions to the hypermultiplet Lagrangian}
The results obtained in the previous subsection allow us to restore 
several leading terms in the hypermultiplet Lagrangian
$L_{\rm H}(G, \vf, {\bar \vf}, \F , {\bar \F}, \S, {\bar \S})$ appearing in (\ref{nact_2}).
Upon elimination of the auxiliary superfields, we find
\bea
\U^I_2 = {\bar \L} \S^I -\hf \G^I{}_{JK}\S^J \S^K 
+\hf \frac{\bar \L}{1+ \L \bar \L } \,
R_{J {\bar L}K}{}^I\,\S^J\S^K
{\bar \S}^{\bar L} 
+ O(\S^4)~,
\label{Gamma2}
\eea
where $O(\S^4)$ denotes all the terms of fourth and higher 
powers in $\S$ and $\bar \S$.

We now project the dynamical superfields to $\cN=1$ superspace 
and consider only the second  Q-supersymmetry transformation \cite{LR,LR-projective}: 
\begin{subequations}
\bea
\d \vf &=& \overline{\e D} \,G~, 
\label{SUSY-tensor1} \\
\d G  &=&  = -\e D\, \vf -\overline{\e D} \, \bar \vf ~,
\label{SUSY-tensor2} \\
\d \F^I &=& \overline{\e D}\,  \S^I~, 
\label{SUSY-hyper1}   \\
\d \S^I &=& -\e D\,  \F^I + \overline{\e D}\,  \U^I_2~,  
\label{SUSY-hyper2} 
\eea 
\end{subequations}
where $\U^I_2$ has to be expressed in terms of the dynamical superfields
as in (\ref{Gamma2}). 
These transformations follow from the relations (\ref{arctic4}), (\ref{arctic5}) and
(\ref{o2n7}), (\ref{o2n8}) by setting $\r^\a =\e^\a ={\rm const}$.
Requiring the hypermultiplet action to possess
this invariance,  and also taking into account the fact that 
$$
L_{\rm H}(G, \vf, {\bar \vf}, \F , {\bar \F}, \S, {\bar \S}) 
= G\, K + O(\S)~,
$$
one can show that
\bea
L_{\rm H}
= G\, K +
\varphi K_I\Sigma^I +\bar\varphi K_{\bar J}\bar \Sigma^{\bar J}
-\frac12 (G+ {\mathbb H}) g_{I\bar J}\Sigma^I\bar\Sigma^{\bar J}
+O(\S^4)~.
\label{leading}
\eea
It can readily be seen that the first three terms generate K\"ahler-invariant 
contributions to the action.
The other terms in $L_{\rm H}$ prove to involve the K\"ahler potential 
ony in the form of the K\"ahler metric $g_{I\bar J}$, the corresponding 
Riemann curvature $R_{I{\bar J}K{\bar L}}$ and its covariant derivatives.

\section{Complex projective space}
\setcounter{equation}{0}

If the K\"ahler potential $K(\F, \bar \F)$  in (\ref{act-hyper})  corresponds to a generic K\"ahler manifold, 
it is not possible to obtain a closed-form expression for the modified geodesic 
equation (\ref{geodesic},\ref{J}) which is equivalent to the auxiliary field equations (\ref{asfem}).
In the non-superconformal case $H=1$,  this equation is known exactly for 
arbitrary Hermitian symmetric spaces \cite{GK,GK2} and is given by eq. 
(\ref{OldGeodesic}). Its extension to the superconformal case is quite  nontrivial, 
due to the presence of an infinite number of curvature-dependent terms in (\ref{J}). 
At the moment, we are not able to derive the equation (\ref{geodesic},\ref{J}) 
even for arbitrary Hermitian symmetric spaces. 
However, below we will work out  explicitly one important example --
the complex projective  space
$\cM= {\mathbb C}P^{n_{\rm H}} = SU(n_{\rm H}+1) / SU(n_{\rm H}) \times U(1)$.  
We believe our consideration for $ {\mathbb C}P^{n_{\rm H}} $  can naturally 
be generalized to the case of arbitrary Hermitian symmetric spaces. 

Using standard inhomogeneous  coordinates for ${\mathbb C}P^{n_{\rm H}} $,
the K\"ahler potential\footnote{Modulo an irrelevant constant, the K\"ahler potential $K (\F, {\bar \F}) $
reduces to (\ref{quadpot})
in the limit  $ r \to \infty$.}
 and the metric are
\be 
K (\F, {\bar \F}) = r^2 \ln \left(1 + \frac{1}{r^2}  
\F^L \overline{\F^L} \right)
~,~~~
g_{I {\bar J}} (\F, \bar \F) =  
\frac{ r^2 \d_{I {J}} }{r^2 + \F^L \overline{\F^L} }
- \frac{  r^2   \overline{\F^I} \F^J  }
{(r^2 + \F^L \overline{\F^L})^2 }~,
\label{s2pot}
\ee
where $I,\bar{J}=1,\dots, n_{\rm H}$ and $r^2={\rm const}$.
The Riemann curvature of  ${\mathbb C}P^{n_{\rm H}}$ is known to be 
\bea
R_{I_1 {\bar  J}_1 I_2 {\bar J}_2}  &:=& 
K_{I_1 {\bar  J}_1 I_2 {\bar J}_2}  
- g_{M \bar N} \G^M_{I_1I_2} {\bar \G}^{\bar N}_{{\bar J}_1{\bar J}_2} 
= -\frac{1}{r^2} \Big\{ g_{I_1 {\bar J}_1 } g_{I_2 {\bar J}_2 }
+g_{I_1 {\bar J}_2 } g_{I_2 {\bar J}_1 } \Big\}~.
\eea
This  implies
\bea
\S^{I_1} {\bar \S}^{ {\bar J}_1 } \S^{I_2}\,
R_{I_1 {\bar J}_1 I_2 {\bar J}_2}  
=-\frac{2}{r^2}  g_{I_2 {\bar J}_2}   \S^{I_2}  |\S|^2~, 
\label{curvature}
\eea
where 
\bea
 |\S|^2 := g_{I \bar J}(\F, {\bar \F})\, \S^{I} {\bar \S}^{ {\bar J} } ~.
\label{quad-inv}
\eea

As before,  let $\U_*(\z) \equiv \U_*( \z; \F, {\bar \F}, \S, \bar \S )$ 
denote the unique solution of the auxiliary field equations (\ref{asfem})
subject to the initial conditions (\ref{geo3}).
Then, the action (\ref{nact}) can be brought to the form
(\ref{nact_2}),
for some Lagrangian $L_{\rm H}$.
Instead of looking directly for $\U_*(\z)$, 
we will try to determine the Lagrangian $L_{\rm H}$
by  making use of considerations based on extended supersymmetry, 
as a generalization of the approaches developed earlier in \cite{AKL2,KN} 
for the non-superconformal case $H=1$.

${}$For  $L_{\rm H}$  we choose an ansatz of the form: 
\bea
L_{\rm H}(G, \vf, {\bar \vf}, \F , {\bar \F}, \S, {\bar \S}) &=& 
G\, K + \varphi K_I\Sigma^I +\bar\varphi K_{\bar J}\bar \Sigma^{\bar J}
+L~, \non \\
L(G, \vf, {\bar \vf}, \F , {\bar \F}, \S, {\bar \S}) &\equiv& L\big( |\S|^2 \big)
=\sum_{n=1}^\infty L_n |\S|^{2n}~, \qquad 
L_n \equiv L_n (G, \vf, \bar \vf )~,  
\label{CPn-lag}
\eea
where the first three terms in the expression for $L_{\rm H}$ agree with 
(\ref{leading}).
The general structure of  $L$ given follows from the fact (\ref{quad-inv}) 
is the only independent $U(n)$-invariant that may be constructed in terms of $\S$s and $\bar \S$s.  
At the moment, we only know that 
\bea
L_1= - \hf (G+{\mathbb H})~.
\label{L-1}
\eea
Our goal is to determine the other Taylor coefficients in (\ref{CPn-lag}), 
$L_2,L_3, \dots,$ using  extended supersymmetry.

Our strategy below will consist in trying to determine $ L\big( |\S|^2 \big)$ 
by requiring the action
\bea 
S_{\rm H}= \int  {\rm d}^4 x \,{\rm d}^4\q 
\,L_{\rm H}(G, \vf, {\bar \vf}, \F , {\bar \F}, \S, {\bar \S}) 
\label{S-hyp}
\eea
to be invariant under the 
second Q-supersymmetry  transformation
(\ref{SUSY-tensor1} -- \ref{SUSY-hyper2}),
with  $\U_2^I$  currently an unknown  function
of the physical superfields which has to be determined.  
We choose the following ansatz for $\U_2^I$: 
\bea
\U^I_2 = -\hf \G^I_{JK} \S^J \S^K 
+\S^I \sum_{n=0}^\infty c_n |\S|^{2n}~, 
 \qquad 
c_n \equiv c_n (G, \vf, \bar \vf )~. 
\label{c}
\eea
At the moment, we only know that 
\bea
c_0=  {\bar  \L} = \frac{2 \bar \vf}{G+{\mathbb H}} ~,
\label{c-0}
\eea
in accordance with (\ref{Gamma2}).
Our goal is to determine the other Taylor coefficients in (\ref{c}), 
$c_1,c_2, \dots,$ using extended supersymmetry.

Let us vary the action with respect to the second 
Q-supersymmetry  transformation
(\ref{SUSY-tensor1} -- \ref{SUSY-hyper2}),
keep the $\bar \e$-dependent terms only, 
and analyze what conditions  are necessary for $S_{\rm H}$ to be invariant. 
The variation $\d_{\bar \e} S_{\rm H}$ involves two types of terms containing  even and  
odd powers of   $\S$s and $\bar \S$s respectively.
The requirement that all even terms vanish can be shown to 
tbe equivalent to the following two conditions:
\bea
\sum_{k=0}^{n-1}  (n-k) c_k L_{n-k} &=&0~, \quad n\geq 2
\label{con1}
\\
\d_{\bar \e} L_n + \sum_{k=1}^n k L_k \overline{\e D} c_{n-k} 
-\frac{1}{n} \sum_{k=1}^n k(n-k) \overline{\e D} \big( L_k c_{n-k} \big) ~
&=& 0~.
\label{con2}
\eea
The requirement that all odd terms vanish can be shown to 
be equivalent to the following condition:
\bea
(n+1) L_{n+1} = \frac{n}{r^2}\,L_n -\vf c_n ~, \quad n\geq 1~.
\label{con3}
\eea

Before continuing the general analysis, let us briefly pause and
make a simple check of  equation (\ref{con2}), by considering the choice $n=1$, that is
\be
\d_{\bar \e} L_1 + L_1 \overline{\e D} c_0 =0~.
\label{con2-1}
\ee
Since 
\be
\d_{\bar \e} (G+{\mathbb H}) = - (G+{\mathbb H}) \overline{\e D} \bar \L ~, 
\ee
the relations (\ref{L-1})  and (\ref{c-0})  imply that (\ref{con2-1}) 
is identically satisfied.

Using the relations (\ref{con1}) and (\ref{con3}), we can derive 
a recursion relation to determine the  coefficients $L_n$. 
It is 
\bea
L_n = \frac{1}{n \mathbb H} \Big\{ 
\sum_{k=1}^{n-2} (n-k)(k+1) L_{n-k}L_{k+1} 
-\frac{1}{r^2} \sum_{k=1}^{n-1} (n-k)k L_{n-k}L_{k}  \Big\} ~, 
\quad n \geq 3~.
\label{con4}
\eea
${}$For $n=2$, only the second term on the right contributes, hence
\bea
 L_2 = -\frac{1}{2 r^2 \mathbb H}  (L_1)^2 = -\frac{1}{8r^2} 
\frac{(G+{\mathbb H})^2}{\mathbb H}~.
\label{con5}
\eea

Making use of eq. (\ref{con4}) allows one to obtain an algebraic equation 
obeyed by 
$$
 L'(x) = \sum_{n=1}^\infty n L_n x^{n-1} ~.
$$
The equation  is 
\bea 
\big(1-{x}/{r^2} \big)  \big[ L'(x) \big]^2 + G L'(x) 
-\frac{1}{4} (   {\mathbb H}^2 - G^2 ) =0~.
\eea
We have to choose the following solution of the quadratic
equation obtained:
\bea
L'(x) = -\hf \frac{G}{1-{x}/{r^2} } 
-\hf   \frac{\sqrt{ {G^2}
+ ({\mathbb H}^2 - G^2 )(1-{x}/{r^2} ) } }{1-{x}/{r^2} } ~, 
\eea
for it possesses the right functional form in the limit  ${\mathbb H} \to G$.
Now, the problem of computing $L(x)$ amounts 
to doing an ordinary integral. The result is as follows:
\bea
L(x) &=&  
- r^2  \Big\{     {\mathbb H} - G \, \ln \big( G+  {\mathbb H}\big) \Big\}  
+r^2 \sqrt{ G^2 + ({\mathbb H}^2 - G^2) ( 1 - |\S|^2/r^2 )} 
\non
\\
&&+ r^2 
G\, \ln \frac{ 1 - |\S|^2/r^2 } 
{ \sqrt{ G^2 + ({\mathbb H}^2 - G^2) ( 1 - |\S|^2/r^2 )} +G }
~.
\label{cotLag3}
\eea
It can be seen  that 
\bea
\lim_ {{\mathbb H} \to G} L(x) =  G \, r^2 \ln\Big( 1 - \frac{x}{r^2} \Big) ~
\eea
which agrees with \cite{GK2,AN,AKL1,AKL2}.

Using the relations (\ref{con3}), we can now compute 
all the coefficients $c_n$, and hence 
the function $c(x)$ appearing 
in (\ref{c}). The latter is 
\bea
c(x):= \sum_{n=0}^\infty c_n x^n = 
2 {\bar \vf} \frac{ 1 - x/r^2 } 
{ \sqrt{ G^2 + ({\mathbb H}^2 - G^2) ( 1 - x/r^2 )} +G }~.
\eea

So far we have determined $L(x)$ and $c(x)$ by using the relations 
(\ref{con1}) and (\ref{con3}). It still remains to be checked that 
eq. (\ref{con2}) is also satisfied. Instead of enjoying such an exercise, 
we choose a different course. 

In accordance with  (\ref{cotLag3}), upon elimination of the auxiliary superfields, 
the hypermultiplet Lagrangian is
\bea
L_{\rm H} &=& G\,K(\Phi, {\bar \Phi})
+ \varphi K_I (\Phi, {\bar \Phi})\Sigma^I 
+\bar\varphi K_{\bar J} (\Phi, {\bar \Phi})\bar \Sigma^{\bar J}
- r^2  \Big\{     {\mathbb H} - G \, \ln \big( G+  {\mathbb H}\big) \Big\}  
\label{L-H}\\
&&+ r^2 \Bigg\{ 
G\, \ln \frac{ 1 - |\S|^2/r^2 } 
{ \sqrt{ G^2 + 4{\bar \vf} \vf ( 1 - |\S|^2/r^2 )} +G }
+  \sqrt{ G^2 + 4{\bar \vf} \vf ( 1 - |\S|^2/r^2 )} 
\Bigg\} ~. ~~~~~
\non
\eea
Consider the second Q-supersymmetry transformation
(\ref{SUSY-tensor1}--\ref{SUSY-hyper2}), 
where 
\bea
\U^I_2 &=&  -\hf \G^I_{JK} \S^J \S^K 
+2\S^I {\bar \vf} \frac{ 1 - |\S|^2/r^2 } 
{ \sqrt{ G^2 + 4{\bar \vf} \vf ( 1 - |\S|^2/r^2 )} +G }~.
\eea
It is an instructive, albeit time consuming,  exercise to check explicitly that the action 
(\ref{S-hyp}), with $L_{\rm H}$ given by (\ref{L-H}), is invariant under 
this transformation. 
This implies that all of the equations (\ref{con2}) are  identically satisfied. 

One can readily check that the action (\ref{S-hyp}) generated 
by the Lagrangian $L_{\rm H} $, eq. (\ref{L-H}), is invariant under 
the $\cN=1$ superconformal transformation (\ref{arctic3}), (\ref{o2n4})
and the shadow chiral rotation (\ref{shadow 2}),
(\ref{shadow 3}), where $n$ should be set to zero for both  transformations.
 We leave it as an exercise for the reader to check that the action is also invariant under 
arbitrary extended superconformal transformations 
(\ref{arctic4}), (\ref{arctic5}) with $n=0$ and (\ref{o2n7}), (\ref{o2n8}).

The hypermultiplet model (\ref{S-hyp}), with $L_{\rm H}$ given by (\ref{L-H}),
possesses a dual formulation obtained by dualizing the complex linear tangent 
variables $\S^I$ and their conjugates
${\bar \S}^{\bar I}$ into chiral one-forms $\J_I$ and their conjugates 
${\bar \J}_{\bar I}$, $ {\bar D}_{\dt \a} \J_I =0$.
As usual, one first replaces the action with a first order one, 
\bea 
S= \int  {\rm d}^4 x \,{\rm d}^4\q 
\,\Big\{ L_{\rm H}(G, \vf, {\bar \vf}, \F , {\bar \F}, \S, {\bar \S})  
+\S^I \J_I +{\bar \S}^{\bar J} {\bar \J}_{\bar J}
\Big\}~,
\eea
where $\S^I $ and ${\bar \S}^{\bar J} $ are chosen to be complex unconstrained.
Next, one eliminates these superfields with the aid of their algebraic 
equations of motions,  ending  up with the dual  Lagrangian:
 \bea
L^{\rm (dual)} _{\rm H} &=& G\,K(\Phi, {\bar \Phi})
- r^2  \Big\{     {\mathbb H} - G \, \ln \big( G+  {\mathbb H}\big) \Big\} 
\label{Calabi} \\
&+& r^2\Big\{   \sqrt{ {\mathbb H}^2 + 4 | \J + \vf \nabla K |^2/r^2 } 
- G\, \ln \Big(   \sqrt{ {\mathbb H}^2 + 4 | \J + \vf \nabla K |^2/r^2 } +G \Big) \Big\}~,~~~
\non
\eea
where
\bea 
 | \J + \vf \nabla K |^2 := g^{I{\bar J}} \Big( \J_I +\vf K_I (\Phi, {\bar \Phi})\Big) 
 \Big( {\bar \J}_{\bar J} +{\bar \vf} K_{\bar J}(\Phi, {\bar \Phi}) \Big) ~. 
 \eea
Under the K\"ahler transformation (\ref{kahl}), the chiral  one-form $\J_I$ changes as
\be
\J_I \quad \longrightarrow \quad \J_I - \vf \,F_I (\F)~,  
\ee
and this transformation is clearly consistent with the chirality of $\J_I$.
In the limit $G=1$ and $\vf =0$, the Lagrangian (\ref{Calabi})  
reduces to the hyperk\"ahler potential for the cotangent bundle of ${\mathbb C}P^{n_{\rm H}}$ 
\cite{Calabi}, see  \cite{LR,AN} and references therein for alternative supersymmetric techniques
to derive the Calabi metric.

To conclude this section, we should mention that the above consideration 
for the complex projective space $\cM=  SU(n_{\rm H}+1) / SU(n_{\rm H}) \times U(1)$ 
can be immediately generalized  
to the non-compact space  $ SU(n_{\rm H},1) / SU(n_{\rm H}) \times U(1)$ 
characterized by the K\"ahler potential 
\be 
K (\F, {\bar \F}) = - r^2 \ln \left(1 - \frac{1}{r^2}  
\F^L \overline{\F^L} \right)~.
\ee
This generalization amounts to replacing everywhere 
$r^2\, \longrightarrow \,-r^2 $.

\section{Discussion} 
\setcounter{equation}{0}

In section 4, we studied the dynamical system ({\ref{nact}) for the case when 
the K\"ahler potential has the form (\ref{s2pot}) and corresponds to 
${\mathbb C}P^{n_{\rm H}}$. Some aspects of this theory are more transparent within its 
dual formulation (\ref{act-dual}) in which the action, 
modulo a trivial rescaling of $\U^I$,  is 
\bea
 S_{\rm dual} &=&  
  \k\, 
 \oint \frac{{\rm d}\z}{2\pi {\rm i} \z} \,  
 \int  {\rm d}^4 x \, {\rm d}^4\q\, \breve{\X} \,\X \,
  \big( 1 + \U^I  \breve{\U}^{\bar I}    \big)^m ~, \qquad m := \frac{r^2}{\k}~.
\label{act-dual2} 
\eea
This formulation is useful to see that the parameter $m$ should be an integer
(compare with \cite{WB}). It is suficient to consider the case of ${\mathbb C}P^1$. 
Then $\U$ is the inhomogeneous complex coordinate in one of the two standard charts 
for ${\mathbb C}P^1 = {\mathbb C} \cup \{\infty\}$, say in the chart  $\mathbb C$.
Let $\U'$ be the complex coordinate in the second chart, ${\mathbb C}^* \cup \{\infty\}$,
with ${\mathbb C}^* :={\mathbb C} - \{0\}$, 
such that the transition function is  $\U' = 1/ \U$.
Of course, the action (\ref{act-dual2}) for ${\mathbb C}P^1$
should be well-defined in both charts. 
In the second chart, it reads 
\bea
S_{\rm dual} &=&
\k\, \oint \frac{{\rm d}\z}{2\pi {\rm i} \z} \,  
 \int  {\rm d}^4 x \, {\rm d}^4\q\, \breve{\X}' \,\X' \,
  \big( 1 + \U' \, \breve{\U}'    \big)^m ~, \qquad \X' = \X \,\U^{m}~.
\label{act-dual3} 
\eea
In order for the compensator $\X'$ to be well-defined on ${\mathbb C}^*$, 
the parameter $m$ should be an integer. 
In the general case of ${\mathbb C}P^{n_{\rm H}}$, 
similar arguments show 
that the varibles $\U^I$ and $\X$ parametrize 
a holomorphic line bundle over ${\mathbb C}P^{n_{\rm H}}$.  

More generally, 
the  arctic variables $\U^I$ and $\X$ in the model (\ref{act-dual}) should parametrize 
a holomorphic line bundle over a K\"ahler-Hodge manifold
 $ \cM^{2n_{\rm H}}_K$ 
with  K\"ahler potential 
\bea 
{\mathbb  K} (\F, \bar \F) = \frac{1}{\k}  K(\F, \bar \F) ~,
\eea
in order for the action to be well-defined.
To justify this claim, it suffices to reiterate the discussion of K\"ahler-Hodge
geometry given in \cite{HitchinKLR} (see also \cite{HLRvUZ} for a recent review).
Let $\o = {\rm i} \pa {\bar \pa} {\mathbb K}$ be the K\"ahler two-form of  $ \cM^{2n_{\rm H}}_K$.
The K\"ahler manifold is   Hodge if $\o / 2\pi \in H^2( \cM^{2n_{\rm H}}_K, {\mathbb Z})$, 
where $H^2( \cM^{2n_{\rm H}}_K, {\mathbb Z})$ denotes the second cohomology group of 
$\cM^{2n_{\rm H}}_K$ with integer coefficients.
Then, one can associate with $\o$  a holomorphic line bundle with connection for which 
$\o$ is the field strength. The K\"ahler potential $\mathbb K$ can be chosen such that 
$h:= {\rm e}^{\mathbb K}$ is a Hermitian fiber metric on the line bundle, 
$|\!| \c |\!|^2 = h \,\c \bar \c$. Given a nowhere vanishing {\it local} section $\c$ of the line bundle, 
the K\"ahler potential can be given, in accordance with \cite{HitchinKLR},  as 
${\mathbb K}= \ln |\!| \c |\!|^2$. 
 This geometric picture extends to the $\cN=2$ supersymmetric case by replacing 
 $\F^I \to \U^I$ and $\c \to \X$. The crucial point is that the action (\ref{act-dual})
 is globally well-defined in spite of the fact that the Lagrangian is given in terms of local 
data.

Our discussion above shows that the dual  formulation  (\ref{act-dual})  
with arctic compensator requires 
K\"ahler-Hodge geometry.  An interesting question is: Can we see the same  geometry 
within the formulation (\ref{nact}) with tensor compensator?
The answer is ``Yes'' provided the action (\ref{nact}) is rewritten in the following equivalent form:
\bea
S [H(\z), \U (\z)] = \k \,
\oint \frac{{\rm d}\z}{2\pi {\rm i} \z} \,  
 \int  {\rm d}^4 x \,{\rm d}^4\q\,
H\, \ln \frac{ {\rm e}^{ {\mathbb  K} (\U, \breve{\U})}\, \X \, \breve{\X}  } {H} ~.
\label{nact5}
\eea
Here $\X (\z) $ is a weight-one arctic multiplet, and $\breve{\X}  (\z)$ its smile-conjugate.
These multiplets are purely gauge degrees of freedom, for  (\ref{nact5})
is invariant under gauge transformations of the form:
\bea
\X  ~ \longrightarrow ~ \X' = {\rm e}^{\r}\, \X~,
\eea
with $\r$ an arbitrary weight-zero arctic multiplet. 
The gauge invariance follows from the identity 
\be
\oint \frac{{\rm d}\z}{2\pi {\rm i} \z} \,  
 \int  {\rm d}^4 x \,{\rm d}^4\q\,
H\, \r =0~.
\ee
Unlike the original action (\ref{nact}), 
its reformulation (\ref{nact5}) is manifestly $\cN=2$ superconformal.\footnote{The action 
 (\ref{nact5}) is the rigid superspace version of a locally supersymmetric action 
 introduced in \cite{K-dual}.}
The  arctic variables $\U^I$ and $\X$ in (\ref{nact5}) parametrize 
the holomorphic line bundle over  $ \cM^{2n_{\rm H}}_K$ introduced earlier.

It should be pointed out that no quantization condition occurs in the case of 
non-conformal $\cN=2$ sigma model (\ref{H=1}). The K\"ahler potential 
in (\ref{H=1}) is required to be real analytic but is otherwise arbitrary. 
The point is that the component Lagrangian can be defined as (compare with \cite{Zumino})
\bea
L_{\rm component} = \frac{1}{16} D^\a {\bar D}^2 D_\a 
 \oint \frac{{\rm d}\z}{2\pi {\rm i}\z}  \,  K \big( \U , \breve{\U}   \big) 
=\frac{1}{16} {\bar D}_{\dt \a} { D}^2 {\bar D}^{\dt \a} 
 \oint \frac{{\rm d}\z}{2\pi {\rm i}\z}  \,  K \big( \U , \breve{\U}   \big) ~,
\eea
and it is manifestly invariant under K\"ahler transformation (\ref{kahl2}).

Let us return to the case  $ \cM^{2n_{\rm H}}_K ={\mathbb C}P^{n_{\rm H}}$ discussed at the beginning
 of this section. As follows from (\ref{act-dual2}), the choice 
\be
r^2 =\k
\ee 
corresponds to a free theory, and this property should also be seen within the original model (\ref{nact}). 
Indeed, for this particular choice of parameters the fourth term in the expression (\ref{L-H}) 
for $L_{\rm H}$ (or the second term in the expression (\ref{Calabi}) 
for the dual Lagrangian $L^{\rm (dual)} _{\rm H}$)
cancels against $\k L_{\rm T}$, with $ L_{\rm T}$ the tensor multiplet Lagrangian, 
eq. (\ref{Lag-tensor}). Now, in the theory with Lagrangian
 \bea
r^2 L_{\rm T} +L^{\rm (dual)} _{\rm H} &=& r^2 \,G \Big\{ \ln \big(1 + \frac{1}{r^2}  
\F^\dagger \F \big)
-  \ln \Big(   \sqrt{ {\mathbb H}^2 + 4 | \J + \vf \nabla K |^2/r^2 } +G \Big) \Big\}
\non
\\
&+& r^2   \sqrt{ {\mathbb H}^2 + 4 | \J + \vf \nabla K |^2/r^2 } ~,
\label{free} 
\eea
one can explicitly dualize 
 the real linear superfield $G$ into
a chiral scalar $\c$ and its conjugate. Modulo a field redefinition, 
the action obtained describes $n_{\rm H}+1$ free hypermultiplets.
Thus, in spite of the fact that the above Lagrangian is nonlinear, 
it generates free dynamics. This is analogous to the situation with the improved tensor 
multiplet (\ref{Lag-tensor}) described in detail in  \cite{LR}. 

In our analysis of the modified geodesic equation in section 3, 
we started with the simplest case of a flat K\"ahler target space characterized 
by the K\"ahler potential (\ref{quadpot}).
This case is actually interesting on its own. 
As mentioned at the end of subsection \ref{quadratic}, 
the polar multiplets $\U^I$ and $\breve{\U}^{\bar I}$ can be dualized into 
real $\cO (2)$ multiplets $\eta^I$ such that the resulting hypermultiplet action 
is given by (\ref{universal}). This action for  $n_{\rm H} =1$ 
provides the projective superspace description  \cite{deWRV}
for the classical universal hypermultiplet \cite{CFG}.
Combining this action with the tensor multiplet sector in (\ref{nact}), 
we obtain a theory of two tensor multiplets with Lagrangian 
\bea
{\mathfrak L}_{\rm UHM} = - \k \,H \ln H - \hf  \frac{\eta^2}{ H} 
\eea
which is (modulo sign)  the projective superspace description \cite{ARV} 
(see also \cite{RoblesLlana:2006ez})
of the one-loop corrected universal hypermultiplet \cite{AMTV}.

There are various interesting problems that can be addressed building on the results of 
this paper. In particular, it is of interest to extend the analysis
for the complex projective space
given in section 4 to the case of arbitrary Hermitian symmetric spaces.
This should include the derivation of  closed-form expressions for  the modified geodesic equation
and the hypermultiplet action $S_{\rm H}$. Such expressions are actually known if 
we set $\vf =0$ and keep only the real linear superfield $G$ of the tensor multiplet $H(\z)$. 
Then, the auxiliary field equations (\ref{asfem}) reduce to those corresponding to the 
non-superconformal model (\ref{H=1}). 
The latter are equivalent, if $\cM$ is Hermitian symmetric, 
 to the geodesic equation (\ref{OldGeodesic}). 
 As to the  hypermultiplet action $S_{\rm H}$, it is obtained 
by inserting $G$ into the integrand of (\ref{act-HS}).
The real challenge, however,  is to extend these simple results, 
corresponding to the special case $H=G$, to the general tensor 
multiplet (\ref{tensor-series}). In the supergravity context, 
the local SU(2) invariance allows one to choose the gauge $H=G$, 
see \cite{KT-M-normal} for a related discussion.
But for the rigid superconformal sigma models under consideration, 
we have at our disposal only rigid SU(2) 
transformations that cannot be used to choose the gauge $\vf=0$ (compare with \cite{RVV}
where such a gauge condition was nevertheless employed).

In conclusion, we mention that our results can be used to study the dynamics 
of  a family of nonlinear  sigma models in $\cN=2$ anti-de Sitter superspace
proposed in \cite{KT-M-AdS}. Such sigma models are described by the action 
(\ref{act-hyper}) in which $H$ is a background tensor multiplet containing 
all the information about the anti-de Sitter supergeometry.
\\

\noindent
{\bf Acknowledgements:}\\
UL and RvU acknowledge the hospitality of the School of Physics at UWA 
where this work was initiated.
We are grateful to Simon Tyler for assistance with Mathematica.
The research   of UL was supported by VR grant 621-2006-3365 and that 
of RvU by Czech ministry of education contract
No. MSM0021622409.

\appendix

\section{N = 2 superconformal transformations and their realization in N = 1 superspace} 
\setcounter{equation}{0}
\label{App1}

General 4D $\cN=2$ superconformal projective multiplets and their superconformal couplings 
were described in detail\footnote{Superconformal $\cO(n)$ multiplets and their couplings were 
also studied in \cite{APSV,IN}, however their analysis was restricted to deriving 
the conditions for invariance under  the SU(2) transformations and dilations.
Unlike the more general  analysis presented in \cite{K-hyper}, 
no results were given in \cite{APSV,IN} for the most interesting 
superconformal projective multiplets -- the polar and tropical multiplets.}
 in  \cite{K-hyper}, building on the earlier equivalent  results in five dimensions
\cite{K-compactified}.
Here we list the $\cN=2$ superconformal transformations of several off-shell 
supermultiplets and their relization in $\cN=1$ superspace following \cite{K-hyper}.

Let  $\U^{[n]} (\z)$ be an arctic weight-$n$ multiplet, 
\be
 \U^{[n]} (\z)= \sum_{n=0}^{\infty} \U_n \z^n~.
 \ee
Its $\cN=2$ superconformal transformation is  as follows:
\bea
\d \U^{[n]} = - \Big(  \x &+&  {\l}^{++} (\z)\,\pa_\z \Big) \U^{[n]}
- n\,\S (\z) \, \U^{[n]}~.
\label{arctic2}
\eea
The smile-conjugate of $\U^{[n]} (\z)$ is the weight-$n$ antarctic multiplet
denoted as $\breve{\U}^{[n]} (\z)$. Its superconformal transformation is 
  \bea
 \d \breve{\U}^{[n]} =  
- \frac{1}{\z^n}\Big(  \x &+&  {\l}^{++} (\z) \,\pa_\z \Big) (\z^n\,\breve{\U}^{[n]} )
-n\,\S (\z) \,\breve{\U}^{[n]} ~.
\label{antarctic2}
\eea
In the case $n=0$, these transformations reduce to (\ref{sctl-U}).

As shown in \cite{K-hyper}, the transformation of $\cN=2$ supermultiplets
associated with the $\cN=2$ superconformal Killing vector $\x$ 
generates three types of transformations
at the level of  $\cN=1$ superfields. 
They are:\\
{\bf 1.} An arbitrary $\cN=1$ superconformal transformation generated by 
\bea
{\bm \x} = {\overline {\bm \x}} = {\bm \x}^a  \pa_a + {\bm \x}^\a D_\a
+ {\bar {\bm \x}}_{\dt \a}  {\bar D}^{\dt \a}
\label{n=1scf1}
\eea  
such that 
\be
[{\bm \x} \;,\; D_\a ] 
= \bm { \o}_\a{}^\b  D_\b +
\Big({\bm \s} - 2 \bar{ {\bm \s}}  \Big) D_\a~,
\label{n=1scf2}
\ee
see \cite{BK} for more detail.
The components of $\bm \x$ and their descendants  $ \bm { \o}_\a{}^\b $ and $\bm \s$
correspond  to the following choice of the $\cN=2$ parameters:
\be
\x \big| ={\bm \x}~, \quad { \o}_\a{}^\b \big|  =\bm { \o}_\a{}^\b ~, 
\quad \s \big|= \bm \s ~,\quad
 {\l}_{\1}{}^{\1} \big|={\bar {\bm \s}}-{\bm \s}~, \quad
 {\l}_{\2}{}^{\1} \big|=0~.
\label{n=1scf3}
\ee
{\bf 2.} An extended  superconformal transformation
generated by 
\bea
\x \big| &=& \r^\a D^{\2}_\a +{\bar \r}_{\dt \a} {\bar D}^{\dt \a}_{\2}~, 
\qquad \x^\a_{\2}  \big|=\r^\a~,
\non \\
{ \o}_\a{}^\b \big|  &=& \s \big|=
 {\l}_{\1}{}^{\1} \big|=0~, \qquad
 {\l}_{\2}{}^{\1} \big|=  {\l}^{\1 \1} \big|=
- \hf D^\a \r_\a~.
\label{esc}
\eea
{\bf 3.} A shadow chiral rotation.
This  is a phase transformation
 of $\q^\a_{\2}$ only, with $\q^\a_{\1}$ kept unchanged, 
and it corresponds to the choice 
\bea
\x \big| =0~, \qquad { \o}_\a{}^\b \big|  =  {\l}_{\2}{}^{\1} \big|=0~, \qquad 
 \s \big| = {\l}_{\1}{}^{\1} \big| = - \bar \s \big | =-\frac{\rm i}{2}\,\a~.
\label{shadow1}
\eea

The spinor parameter $\r^\a$ in (\ref{esc}) can be shown to obey the equations
\bea
{\bar D}_{\dt \a} \r^\b =0~, \qquad D^{(\a}\r^{\b )}=0~,
\eea
and the latter imply 
\be
\pa^{{\dt \a} (\a} \r^{\b )} = D^2 \r^\b =0~.
\ee
There are several ordinary (component) transformations generated by 
the chiral spinor $\r^\a$ in (\ref{esc}): (i) second Q-supersymmetry 
transformation $(\e^\a$); (ii) off--diagonal SU(2)-transformation 
($  {f} =  {\l}^{\1 \1}|_{\q=0} $); (iii) second S-supersymmetry transformation 
(${\bar \eta}_{\dt \a}$). They emerge as follows: 
\be
\r^\a(x_{(+)}, \q) = \e^\a
+ f \q^\a - {\rm i} \,{\bar \eta}_{\dt \a}\, x^{{\dt \a}\a}_{(+)}~,
\ee
 with $x_{(+)}^a$ the chiral extension of $x^a$.

Consider the arctic weight-$n$ multiplet $\U^{[n]} (\z)$. 
Its $\cN=2$ superconformal transformation law (\ref{arctic2})
generates the following $\cN=1$ variations
of the component superfields:\\
{\bf 1.}  the $\cN=1$ superconformal transformation
\bea
\d \U_k = -{\bm \x} \U_k - 2k(\bar{\bm \s} - {\bm \s})\U_k -2n {\bm \s}\U_k~;
\label{arctic3} 
\eea
{\bf 2.} the extended  superconformal transformation
\begin{subequations}
\bea
\d \U_0 &=& {\bar \r}_{\dt \a} {\bar D}^{\dt \a} \U_1 
+\hf \big( {\bar D}_{\dt \a} {\bar \r}^{\dt \a} \big) \U_1 ~,
\label{arctic4}  \\
\d \U_1 &=&-\r^\a D_\a \U_0 
+  {\bar D}_{\dt \a}\big( {\bar \r}^{\dt \a}  \U_2 \big)
-\frac{n}{2}\big(D^\a\r_\a\big) \U_0 ~, 
\label{arctic5} \\
\d \U_k &=&-\r^\a D_\a \U_{k-1} 
+{\bar \r}_{\dt \a} {\bar D}^{\dt \a} \U_{k+1} \non \\
&&
+\hf (k-n-1)\big(D^\a\r_\a\big) \U_{k-1} 
+\hf (k+1) \big( {\bar D}_{\dt \a} {\bar \r}^{\dt \a} \big) \U_{k+1}
~, \qquad k>1~; ~~~~
\label{arctic6} 
\eea
\end{subequations}
{\bf 3.} the shadow chiral rotation
\be
\d \U_k = {\rm i}\,\a (k-\frac{n}{2}) \U_k~.
\label{shadow 2}
\ee
Choosing $n=0$ in the above relations, one obtains the transformations 
of the dynamical superfields $\F:= \U_0$ and  $\S:= \U_1$ 
of the weight-zero arctic multiplet $\U$.

Consider the tensor multiplet $H(\z)$, eq. (\ref{tensor-series}). 
Its $\cN=2$ superconformal transformation law (\ref{sctl-H})
generates the following $\cN=1$ transformations
of the component superfields:\\
{\bf 1.}  the $\cN=1$ superconformal transformation
\bea
\d \vf = -{\bm \x} \vf  -4{\bm \s} \vf~, \qquad 
\d G = -{\bm \x} G - 2( {\bm \s} + \bar{\bm \s} )G~;
\label{o2n4} 
\eea
{\bf 2.} the extended  superconformal transformation
\begin{subequations}
\bea
\d \vf &=& {\bar \r}_{\dt \a} {\bar D}^{\dt \a} G
+\hf \big( {\bar D}_{\dt \a} {\bar \r}^{\dt \a} \big) G ~, 
\label{o2n7} \\
\d G &=&-\r^\a D_\a \vf 
-{\bar \r}_{\dt \a} {\bar D}^{\dt \a} \bar{\vf} 
-\Big((D^\a\r_\a) \vf 
+\big( {\bar D}_{\dt \a} {\bar \r}^{\dt \a} \big) 
\bar{\vf} \Big)~;~~~~~~
\label{o2n8} 
\eea
\end{subequations}
{\bf 3.} the shadow chiral rotation
\be
\d \vf = - {\rm i}\,\a \vf~, \qquad \d G=0~.
\label{shadow 3}
\ee

\section{Supersymmetric  sigma models on tangent bundles of Hermitian symmetric spaces} 
\setcounter{equation}{0}

This appendix contains a summary of several results obtained in a series of papers
\cite{GK,GK2,AN,AKL1,AKL2,KN} devoted to the study of $\cN=2$ supersymmetric 
sigma models of the form (\ref{H=1}), where $K(\F,  \bar \F )$ is the K\"ahler potential 
of a Hermitian symmetric space,  and therefore
 the corresponding  curvature tensor is covariantly constant,
\be
\nabla_L  R_{I_1 {\bar  J}_1 I_2 {\bar J}_2}
= {\bar \nabla}_{\bar L} R_{I_1 {\bar  J}_1 I_2 {\bar J}_2} =0~.
\label{covar-const}
\ee
In such a model, the auxiliary field equations are equivalent 
to the geodesic equation with complex evolution parameter \cite{GK,GK2}
\bea
\frac{ {\rm d}^2 \U^I (\z) }{ {\rm d} \z^2 } + 
\G^I_{~JK} \Big( \U (\z), \bar{\F} \Big)\,
\frac{ {\rm d} \U^J (\z) }{ {\rm d} \z } \,
\frac{ {\rm d} \U^K (\z) }{ {\rm d} \z }  =0~.
\label{OldGeodesic}
\eea
Upon elimination of the auxiliary superfields, the action (\ref{H=1}) 
can be shown to take the form  \cite{KN}:
\bea
S [\U_*(\z)]&=&    
 \int  {\rm d}^4 x \, {\rm d}^4\q\,  \Big\{ K \big( \F, \bar \F  \big) 
- \hf {\bm \S}^{\rm T} {\bm g} \,
\frac{ \ln \big( {\mathbbm 1} + {\bm R}_{\S,\bar \S}\big)}{\bm R_{\S,\bar \S}}
\, {\bm \S}\Big\}~, \quad 
{\bm \S} :=\left(
\begin{array}{c}
\S^I\\
{\bar \S}^{\bar I} 
\end{array}
\right) ~, ~~~~~
\label{act-HS}
\eea
where 
\bea
{\bm R}_{\S,\bar \S}
:=\left(
\begin{array}{cc}
0 & (R_\S)^I{}_{\bar J}\\
(R_{\bar \S})^{\bar I}{}_J &0 
\end{array}
\right)~, 
\quad (R_\S)^I{}_{\bar J}:=\hf R_K{}^I{}_{L \bar J}\, \S^K \S^L~, 
\quad (R_{\bar \S})^{\bar I}{}_J := \overline{(R_\S)^I{}_{\bar J}}~,~~~~
\label{R-Sigma}
\eea
and
\bea
{\bm g}
:=\left(
\begin{array}{cc}
0 & g_{I \bar J}\\
g_{{\bar I}J} &0 
\end{array}
\right)~.
\eea
Here $\U_*(\z)$ denotes the unique solution of equation (\ref{OldGeodesic})
under the initial conditions (\ref{geo3}). 
A different universal representation for the action $S [\U_*(\z)]$ can be found in 
\cite{AKL2}.

\small{

}

\end{document}